# Physical Design and Monte Carlo Simulations of a Space Radiation Detector onboard the SJ-10 satellite

LIU Ya-qing(刘雅清)[1,2] WANG Huan-yu(王焕玉)[1;1)] CUI Xing-zhu(崔兴柱)[1] PENG Wen-xi(彭文溪)[1] FAN Rui-rui (樊瑞睿)[1] LIANG Xiao-hua(梁晓华)[1] GAO Ming(高旻)[1,2] ZHANG Yun-long(张云龙)[1,3] ZHANG Cheng-mo(张承模)[1] ZHANG Jia-yu(张家宇)[1] YANG Jia-wei(杨家卫)[1] WANG Jin-zhou(汪锦州)[1] ZHANG Fei(张飞)[1] DONG Yi-fan(董亦凡)[1,2] GUO Dong-ya(郭东亚)[1,2] ZHOU Da-wei(周大卫)[1,2]

[1] Institute of High Energy Physics, Chinese Academy of Sciences, Beijing 100049, China
[2] University of Chinese Academy of Sciences, Beijing 100049, China
[3] Jilin University, Changchun 130012, China

**Abstract:** A radiation gene box (RGB) onboard the SJ-10 satellite is a device carrying mice and drosophila cells to determine the biological effects of space radiation environment. The shielded fluxes of different radioactive sources were calculated and the linear energy transfers of γ-rays, electrons, protons and α-particles in tissue were acquired using A-150 tissue-equivalent plastic. Then, a conceptual model of a space radiation instrument employing three semiconductor sub-detectors for deriving the charged and uncharged radiation environment of the RGB was designed. The energy depositions in the three sub-detectors were classified into fifteen channels (bins) in an algorithm derived from the Monte Carlo method. The physical feasibility of the conceptual instrument was also verified by Monte Carlo simulations.

**Key words:** particle detecting; space radiation; detector design; Monte Carlo simulations

---

[1)] wanghy@ihep.ac.cn

# 1 Introduction

Due to the absence of the earth's atmosphere and magnetospheric shielding, the intensity of natural radiation fluxes greatly increases during space exploration missions. Thus, the radiation hazards to astronauts should be prudently considered. Space radiation can damage the spleen, bones and lymph and may even lead to haemorrhage or cancer. Among these potential effects, more attention is generally paid to the destruction of the genome (DNA), which may lead to pathological states in tissues, organs and systems.

In the past forty years, space radiation studies [1] have focused on the detections and radiation experiments of the space cell incubators, such as radiation detectors and culture devices for drosophila. In this study, a radiation gene box (RGB) was designed to study the status of the radiation-sensitive mice and drosophila cells before and after space travel and to monitor the radiation environment of the cells. The RGB was positioned inside the re-entry capsule of the SJ-10 satellite. A conceptual model of a radiation instrument employing three semiconductor sub-detectors was designed to fulfil the environmental monitoring requirements.

Two constraints were considered for the radiation instruments. First, the size of the detector system must be compact enough to provide adequate room for maintaining a normal life in the mice and drosophila cells. Second, the detection accuracy should be sufficient to detect and identify various particles over a wide energy range.

In this paper, on the basis of the two key points, a space radiation detector for the SJ-10 satellite is designed and its feasibility is validated by the detector's simulation. Moreover, the linear energy transfers (LETs) in tissue are derived for various particles with different energies and an algorithm for identifying these particles is proposed in order to provide reliable radiation environment data for the biological experimental apparatus.

# 2 Analysis of space radiation

Radiation dose is an important factor in space radiation. It can be calculated by the following equation [2].

$$D = 1.60 \times 10^{-10} \times \phi \times \left(\frac{dE}{dx}\right) \quad , (1)$$

where $D$ is the dose in $Gy$, $\phi$ is the particle fluence in particles/cm$^2$ and $\frac{dE}{dx}$ is the LET in $MeV/(g/cm^2)$.

The SJ-10 is a low earth orbit (LEO) satellite, which moves in an elliptical orbit around the Earth at an altitude between 180 km and 460 km. In a LEO satellite, the dominant particles are the protons and electrons. They are the primary ionizing radiation sources and mainly emitted by galactic cosmic rays (GCRs), the protons and electrons trapped in the earth's magnetic field and the particles from solar events. However, around and within the space radiation detector, there were other types of radiation particles and a more complex radiation field. These included not only the primary ionizing radiation in the LEO satellite but also the secondary radiation in the biological objects and the shielding materials of the spacecraft.

As the primary radiation source, the GCR spectrum is composed of 98% heavier ions (baryon component) and 2% electrons and positrons (lepton component). The baryon component contained 87% protons, 12% α-particles and 1% heavy ions [3].

To evaluate the orbital cosmic rays, the galactic cosmic ray model (ISO-15390 standard model) was used and the fluxes between different baryon components were compared. The results showed that: (i) The proton flux was much greater than the α-particle flux. (ii) The integral flux of the α-particles was approximately 10 times larger than the integral flux of other ion species. (iii) Due to the shielding provided by the earth's atmosphere and magnetosphere, the differential flux of the α-particles below 30 MeV was less than 0.001 $cm^{-2}s^{-1}$, and the flux of the α-particles at low energies was rather small in the LEO satellite. Subsequently, the shielding effect by 7.62 mm Al on the α-particle differential energy spectra was evaluated with the NOVICE code. The results indicated that because the secondary particles were produced by the energetic particles, the Al shielding could not reduce the flux of the α-particles to less than 30 MeV [4].

In addition to the GCRs, the fluxes of the protons and electrons trapped in the earth's magnetic field were also estimated. As shown in Figures 1 and 2, the orbitally-shielded fluxes of the electrons and protons were estimated using the AE-8 [5] and AP-8 [6] models, respectively, with 3 mm Al shielding. Figure 1 illustrates the flux of electrons above 0.1 MeV. More specifically, the flux of electrons at 0.5 MeV and 2 MeV was approximately 24 $cm^{-2}s^{-1}$ and 1 $cm^{-2}s^{-1}$, respectively, and the flux above 10 MeV was less than 0.001 $cm^{-2}s^{-1}$. As shown in Figure 2, the differential flux of 5–200 MeV protons was much larger than that of the protons in other energy bands.

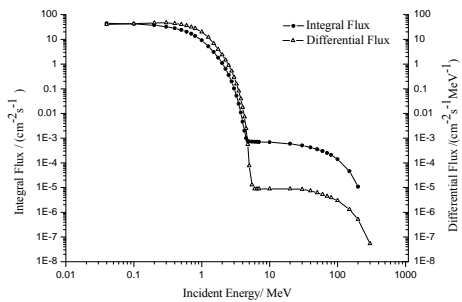

Fig. 1 The orbitally-shielded flux of electrons (the shielding material is 3.705 mm Al, which is equivalent to a mass thickness of 1.0 g/cm$^2$)

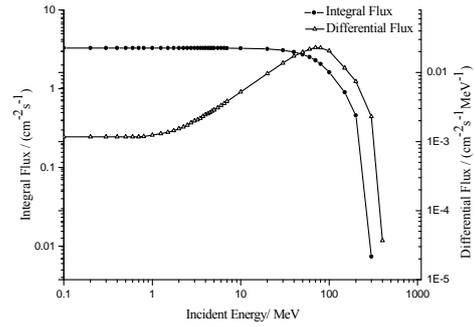

Fig. 2 The orbitally-shielded flux of the protons (the shielding material is 3.705 mm Al, which is equivalent to a mass thickness of 1.0 g/cm$^2$)

In addition, because of the lower solar activity level during the SJ-10 mission, solar energetic particles were not considered as a primary radiation source to the SJ-10 space radiation detector.

After evaluating the fluxes of the primary radiation sources, the fluxes of the secondary radiation sources were analyzed. With respect to the secondary radiation, the electron flux around and within the SJ-10 space radiation detector should increase and approximately 10 keV–20 MeV γ-rays may be produced by the interaction of the energetic particles. The flux of albedo γ-rays varied as an inverse power law of cutoff rigidity with an exponent of approximately 1.9–1.65 [7]. Our calculations showed that the flux of 0.1–2 MeV γ-rays drops from 40 $cm^{-2}s^{-1}$ to approximately 0.1 $cm^{-2}s^{-1}$.

On the basis of the above flux analysis, it could be concluded that the four kinds of large flux particles, i.e. protons, electrons, α-particles and γ-rays, exist in the radiation field of the space detector. Thus, the detector should be designed to identify these four kinds of particles. The detection energy ranges of these particles should be as follows: The α-particles have a lower limit of 30 MeV by the geomagnetic cutoff. The electron energy range would normally be 0.1–10 MeV; however, because of the limits in the detection precision, this range can only be 0.5–10 MeV. Moreover, the detection energy range can be 5–200 MeV for the protons and 0.1–2 MeV for the γ-rays.

For the quantitative understanding of the radiative effects on the human body, it is necessary to know the LET. The ESTAR, PSTAR and ASTAR databases, on the basis of the methods described in the International Commission on Radiological Protection (ICRU) reports 37 and 49 [8], were used to calculate the LET of electrons, protons and α-particles in A-150 tissue-equivalent plastic [9]. The tissue-equivalent material contained 10.13% H, 77.55% C, 3.50% N, 5.23% O, 1.74% F and 1.84% Ca by weight, with a density of 1.127 g/cm$^3$. Figures 3–5 show the LET tables, with the data sourced from the ESTAR, PSTAR and ASTAR databases.

As shown in Figure 3, with an increase in the electron incident energy, the electron LET decreased initially and then reached an equilibrium value. When the electron incident energy was above 2 MeV, the LET increased slowly due to the bremsstrahlung emission. Therefore, according to these trends, the low-energy electrons were divided into three channels : 0.5–1 MeV, 1–2 MeV and 2–10 MeV.

As shown in Figure 4, the relatively low-energy protons (19 MeV) and high-energy α-particles (75 MeV/nuc) had the same LET (approximately 28 MeV/(g/cm2)). Therefore, the protons had a longer range in tissue than the α-particles with the same incident energy. However, with respect to the shielded flux and the secondary protons that may be produced within the tissue by either a slowing down or a nuclear interaction, more channels should be chosen for the protons. Furthermore, because the high-energy α-particles have a wide energy range but a small flux, 300 MeV was chosen as the maximum measurable energy in order to control the size of the space detector.

Next, the mass energy-absorption coefficient of γ-rays (gamma LET) is calculated for the A-150 tissue-equivalent plastic [8]. As shown in Figure 5, the mass energy-absorption coefficient of γ-rays (0.1–2 MeV) showed little or no difference with an increase in the γ-rays incident energy. Thus, only one channel was selected for detecting the γ-rays.

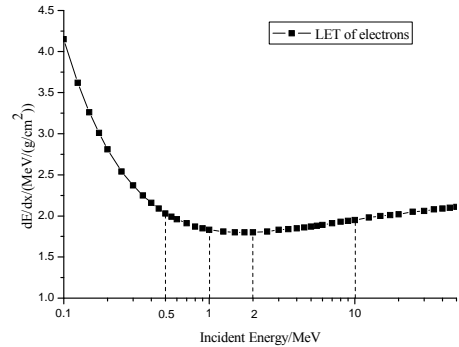

Fig. 3 LET of electrons in A-150 tissue-equivalent plastic

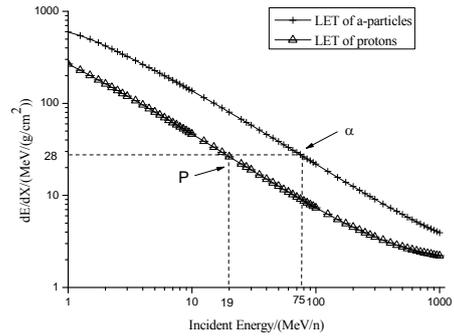

Fig. 4 LET of α-particles and protons in A-150 tissue-equivalent plastic

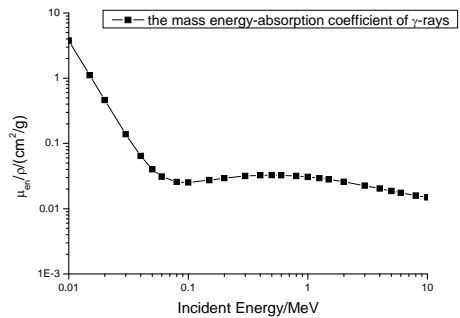

Fig. 5 The mass energy-absorption coefficient of γ-rays in A-150 tissue-equivalent plastic.

Table 1 shows the particles and their measurable energy ranges in the space radiation detector.

Table 1. Detection Ranges

| particle | Energy range | Channel number |
|---|---|---|
| γ–ray | 0.1–2 MeV | 1 |
| electron | 0.5–10 MeV | 3 |
| proton | 5–200 MeV | 8* |
| α-particle | 30–300 MeV | 3* |

*the channel number of the protons and the α-particles is determined after the detector simulation.

## 3 Conceptual design of the detector

Using the Bethe formula [10], the loss of charged particle ionization energy can be determined from the following equation:

$$(-dE/dx)_{ion} \times E \propto Mz^2, \quad (2)$$

where $(dE/dx)_{ion}$ is the energy loss per unit distance, $M$ is the incident particle mass, $z$ is the incident particle charge and $E$ is the incident particle energy.

$(dE/dx)_{ion} \times E$ is proportional to $M$ and $Z^2$. Therefore, by measuring $dE \times E$, the charged particles can be distinguished and the energy ranges can be determined. So the space detector system should include both a $dE$ sub-detector and an $E$ sub-detector to measure $dE$ and $E$, respectively.

When γ-rays passed through the detector, the interaction between the γ-rays and the detector could be due to the Compton effect, the electron pair effect or the photoelectric effect. The energy deposited is proportional to the atomic number and the thickness of the detector. However, compared to that of charged particles, the energy deposition of the γ-rays was much lower. Thus, to improve the sensitivity response to the γ-rays, cadmium zinc telluride (CZT) was selected as the detector material. Its high sensitivity towards the γ-rays allowed the monitoring of the γ-rays from the secondary radiation.

On the basis of the above analysis, three semiconductor sub-detectors were selected to detect the particles based on their compact size and superior resolution. Figure 6 illustrates the structure of the space radiation detector system. The three sub-detectors were surrounded by a 5 mm Al protective enclosure. Moreover, a polyimide window was installed at the front of the detector system to exclude the visible light. The field of view (FOV) of the first detector was 30°.

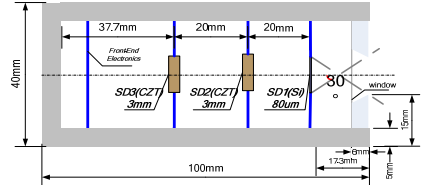

Fig. 6 Schematic of the space radiation detector system.

Sub-detector one (SD1), shown on the right side in Fig. 6, is a 10 mm × 10 mm × 80 μm Si-PIN detector designed to measure the partial energy deposition $dE$ of charged particles within the re-entry capsule. This thin sub-detector satisfied our $dE$ detection requirements. Sub-detector two (SD2), the middle detector in the figure, is a 10 mm × 10 mm × 3 mm CZT detector designed to measure the total energy deposition $E$ of the relatively low-energy charged particles and the γ-rays. Sub-detector three (SD3), the detector on the left side of the figure, is a 10 mm × 10 mm × 3 mm CZT detector designed to detect the relatively high-energy charged particles that can pass through SD1 and SD2. Moreover, SD3 also had some sensitivity to the γ-rays that passed through SD1 and SD2.

## 4 Verification by the simulation

The energy depositions in the three sub-detectors were simulated using the Monte Carlo tool Geant4, which is is a platform for the simulation of the passage of particles through matter. Each simulation used one hundred thousand particles. One hundred thousand was less than the minimum total number of these four kinds of particles on board. The simulation contained the Compton effect, the electron pair effect and the photoelectric effect of the γ-rays and the collisions, ionization and bremsstrahlung emission of charged particles.

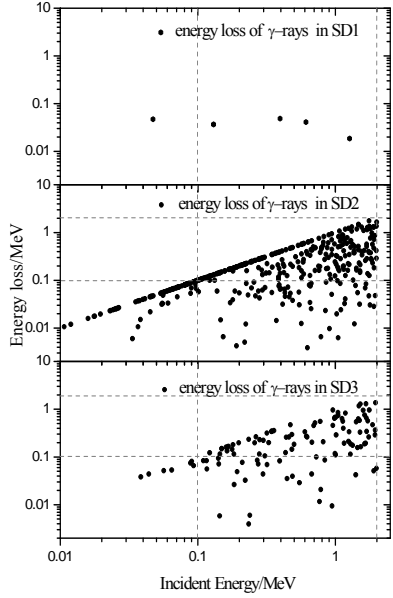

Fig. 7 simulation data obtained from 0-2MeV γ-rays fluence. Each panel shows a scatter plot of the energy depositions in three detectors against the incident energy. The thresholds are marked with the dotted lines.

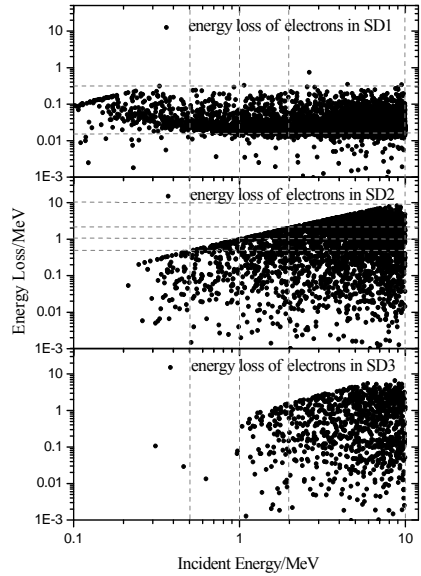

Fig. 8 simulation data obtained from 0-10MeV electrons fluence. One channel is selected for electrons which pass through the SD1.

As shown in Figure 7, the γ-ray energy was mainly deposited in SD2 and SD3 and scarcely deposited in SD1, which was helpful in identifying the γ-rays. Figure 8 shows that the electrons deposited their energies in all the three detectors because the scatter cross-section of the low-energy electrons was large. Furthermore, the electron energy deposition was related to the incident electron energy.

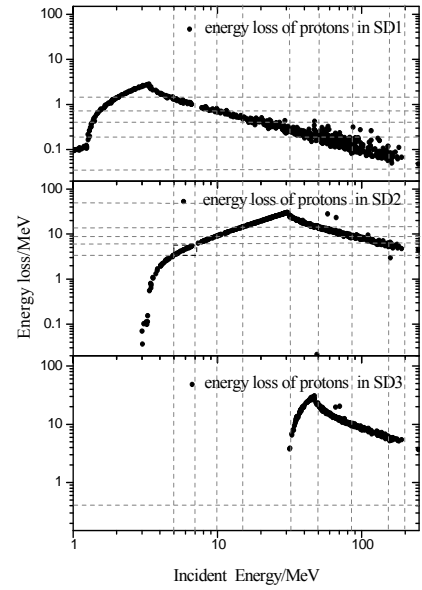

Fig. 9 simulation data obtained from 0-200 MeV protons

Figures 9 and 10 show the energy depositions of the protons and α-particles. Their energy depositions in the different sub-detectors were highly correlated with their incident energies. When they passed through a sub-detector, because of their large incident energies, the energy depositions in the sub-detector correspondingly decreased. As given by Eq. (2), the charge and mass of the α-particles are higher than those of the protons. Thus, their energy depositions were higher. This information was helpful for distinguishing the α-particles from the protons. Therefore, the protons that stopped in SD2 were divided into four channels, those passing through SD2 were divided into another four channels, and the α-particles were divided into just three channels due to their small quantity.

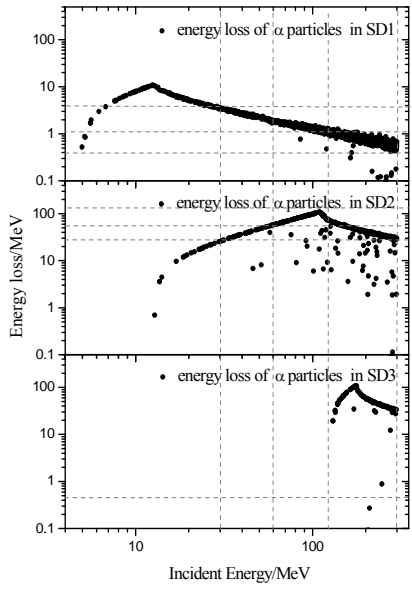

Fig. 10 Simulation data obtained from 0-300MeV α- particles fluence.

On the basis of the energy depositions in the different sub-detectors, the channel range could be determined. As shown in Figure 11, different kinds of particles were in different zones so that they could be identified Thus a detection algorithm was established, as shown in Table 2. A particle can be identified if its energy loss in the three detectors coincides with the discrimination algorithm table. The discrimination algorithm was running on the FPGA in form of lookup table and it was fulfiled after the analogy signals were converted to digital.

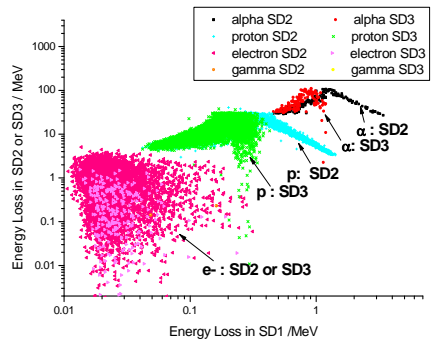

(a)

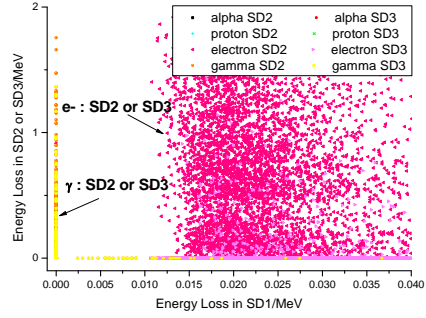

(b)

Fig. 11 Particle identification plots from the simulation. (a) the α-particle, proton and electron simulations, (b) the γ-ray and electron simulations.

Table 2. Thresholds in the different sub-detectors.

| channel | $dE_1$ in SD1/Mev | $dE_2$ in SD2/MeV | $dE_3$ in SD3/MeV | Discrimination algorithm | Incident E /MeV |
|---|---|---|---|---|---|
| gamma1 | >0.016 | 0.1-2 | 0.1–2 | $\overline{dE_1} \cdot dE_2 \cdot dE_3$ | 0.1–2 |
| electron1 | 0.016-0.30 | 0.45-1.0 | - | $dE_1 \cdot dE_2$ | 0.5–1.0 |
| electron2 | 0.016-0.30 | 1.0-2.0 | - | $dE_1 \cdot dE_2$ | 1.0–2.0 |
| electron3 | 0.016-0.30 | 2.0-10.0 | - | $dE_1 \cdot dE_2$ | 2–10 |
| proton1 | 0.4-1.5 | 3.3-5.7 | >0.5 | $dE_1 \cdot dE_2 \cdot \overline{dE_3}$ | 5–7 |
| proton2 | 0.4-1.5 | 5.7-9.1 | >0.5 | $dE_1 \cdot dE_2 \cdot \overline{dE_3}$ | 7–10 |
| proton3 | 0.4-1.5 | 9.1-14.5 | >0.5 | $dE_1 \cdot dE_2 \cdot \overline{dE_3}$ | 10–15 |
| proton4 | 0.2-0.7 | 14.5-55 | >0.5 | $dE_1 \cdot dE_2 \cdot \overline{dE_3}$ | 15–35 |
| proton5 | 0.04-0.4 | 14.5-55 | >0.5 | $dE_1 \cdot dE_2 \cdot dE_3$ | 35–50 |
| proton6 | 0.04-0.4 | 9.1-14.5 | >0.5 | $dE_1 \cdot dE_2 \cdot dE_3$ | 50–85 |
| proton7 | 0.04-0.4 | 5.7-9.1 | >0.5 | $dE_1 \cdot dE_2 \cdot dE_3$ | 85–150 |
| proton8 | 0.04-0.4 | 3.3-5.7 | >0.5 | $dE_1 \cdot dE_2 \cdot dE_3$ | 150–200 |
| Alpha1 | 1-4 | 25-55 | >0.5 | $dE_1 \cdot dE_2 \cdot \overline{dE_3}$ | 30–60 |
| Alpha2 | 1-4 | 55-120 | >0.5 | $dE_1 \cdot dE_2 \cdot \overline{dE_3}$ | 60–130 |
| Alpha3 | 0.4-1 | 25-120 | >0.5 | $dE_1 \cdot dE_2 \cdot dE_3$ | 130–300 |

## 5 Summary

The purpose of designing a space radiation detector was to detect space particles that may cause damage to the DNA. In this study, we analyzed the types of particles within the space radiation environment. Meanwhile, on the basis of the satellite orbit, we estimated the associated particle fluxes. We then calculated the LETs of these particles in tissue-equivalent plastic to find the energy ranges of interest. Thus, we were able to acquire the types and energy ranges of the detected particles. On the basis of the analysis of the interaction between the particles and matter, we designed a telescope-structure detector system containing one silicon detector and two CZT detectors for identifying these particles. We simulated the telescope-structure detector system and the particle energy deposition within the space radiation detector system The Monte Carlo simulation results proved that the system design requirements were satisfied. Thus, we successfully identified the electrons, protons, α-particles and γ-rays using a rather compact structure design.